\documentclass[twocolumn,a4paper,preprintnumbers,amsmath,amssymb,nofootinbib,floatfix]{revtex4}

\usepackage{color}
\usepackage{graphicx}
\usepackage{dcolumn}
\usepackage{bm}
\usepackage{latexsym}
\usepackage{epsfig}
\usepackage{rotating}
\usepackage{bm}

\newcommand{\be}{\begin{equation}}
\newcommand{\ee}{\end{equation}}
\newcommand{\beqq}{\setlength\arraycolsep{2pt}\begin{eqnarray}}
\newcommand{\eeqq}{\vspace{0cm} \end{eqnarray}}
\newcommand{\bea}{\begin{eqnarray}}
\newcommand{\eea}{\end{eqnarray}}

\begin{document}


\title{X-ray surface brightness observations of galaxy clusters, cosmic opacity and the limits on the matter density parameter}

\author{R. F. L. Holanda$^{1,2}$} \email{holandarfl@gmail.com}
\author{Kamilla V. R. A. Silva$^{2}$} \email{kamillaveronika@gmail.com}
\author{V. C. Busti$^{3,4}$} \email{busti@sas.upenn.edu}

\affiliation{ \\$^1$Departamento de F\'{\i}sica, Universidade Federal de Sergipe, 49100-000, Aracaju - SE, Brazil,
\\ $^2$Departamento de F\'{\i}sica, Universidade Federal de Campina Grande, 58429-900, Campina Grande - PB, Brazil,\\$^3$Department of Physics and Astronomy, University of Pennsylvania, Philadelphia, PA 19104, USA, \\
$^4$Departamento de F\'{\i}sica Matem\'atica, Universidade de S\~ao Paulo, Rua do Mat\~ao 1371, S\~ao Paulo - SP, 05508-090, Brazil}



\begin{abstract}

In this work, we use two gas mass fraction samples  of galaxy clusters obtained from their X-ray surface brightness observations  jointly with recent $H(z)$ data in a flat $\Lambda$CDM framework to impose limits on cosmic opacity.{   It is assumed that the galaxy clusters are in 
hydrostatic equilibrium and their gas mass fraction measurement is constant with redshift.} We show that the current limits on the  matter density parameter obtained from X-ray gas mass fraction test are strongly dependent on the  cosmic transparency assumption even for a flat scenario. Our results are consistent with a transparent universe within $1\sigma$ c.l. in full agreement with other analyses which used type Ia supernovae, gamma ray burst and $H(z)$ data.

\end{abstract}

\maketitle


\section{Introduction}

 From a general point of view,  cosmic opacity can be an important systematic error source in several astronomical observations. By considering type Ia supernova (SNe Ia) observations, for instance, there are four different sources of opacity by dust { {absorption}}: the Milky Way, the hosting galaxy, intervening galaxies, and the Intergalactic Medium \cite{combes,conley,menard,imara,McKinnon}. In this context, the  {approach} of Ref.\cite{lima2011} considered SNe Ia data and two different scenarios with cosmic absorption and the main conclusion was that  the description of an accelerating Universe powered by dark energy\footnote{For details please refer to Ref.\cite{weinberg}.} or some alternative gravity theory only must be invoked if  the cosmic opacity is fully negligible. Similar studies also can be found in Refs.\cite{basset,chen1,chen2,Li2013}. The Ref.\cite{xie} investigated the luminosity and redshift dependence of the quasar continuum and suggested that  the reddening observed could come from cosmic dust extinction. Infrared surveys { {can also be affected}} by a population of dust grains \cite{fymbo}. By considering the cosmic microwave background radiation, the results of Ref.\cite{inara} showed that whereas the dust emission in galaxies could be taken out, the intergalactic dust emission is diffuse and cannot be removed easily from the maps. There is also a more exotic possibility as opacity source, namely,  photons turning into unobserved particles beyond the standard model due to interaction with extragalactic magnetic fields (please refer to \cite{avg2009,avg2010,jac2010,tiw} for details).  

Other cosmic opacity tests have been  performed: the approaches of Refs.\cite{HCA2013,LAZ2015} used current measurements of the expansion rate $H(z)$ and SNe Ia data to impose cosmological model-independent constraints on cosmic opacity. As a result, a fully transparent universe is in agreement with the data considered (see also Ref.\cite{avg2009,avg2010} for analyses in a flat $\Lambda$CDM framework).  In order to explore a possible presence of an opacity at higher redshifts $(z > 2)$, the Ref.\cite{HB2014} considered  $H(z)$ data and luminosity distances of gamma-ray bursts in  the $\Lambda$CDM and $\omega$CDM flat models. More recently, the Ref.\cite{FHA}  used 32 old passive galaxies and SNe Ia data to obtain cosmological model-independent constraints on cosmic opacity. No significant opacity was found  in these studies although the results do not completely rule out the presence of some dimming source and  additional tests are still required. This is exactly the subject of the present paper, where  X-ray gas mass fraction samples of galaxy clusters jointly with  recent $H(z)$ data will be used. It is worth emphasizing that X-ray astronomy provides an unique opportunity to detect opacity sources that may be missed by traditional detection methods, such as those using the dust reddening of background quasars by foreground galaxies and associated large scale structure \cite{wright,lia,menardb}.

The gas mass fraction is defined as $f =M_{gas}/M_{Tot}$ \cite{Sasaki96}, where $M_{Tot}$ is the total mass and it can be obtained via hydrostatic equilibrium assumption while  $M_{gas}$ (gas mass) is obtained by  integrating a gas density model (see next section for details) and using X-ray or Sunyaev-Zel'dovich effect observations. By using hot, massive and relaxed galaxy clusters as laboratories, gas mass fraction samples have been compiled and used to  constrain cosmological parameters, mainly the matter density parameter, $\Omega_M$ (see \cite{allen2002,lima2003,roque2006,allen2008,ettori2009,mantz2014} for several analyses).  The gas mass fraction as a cosmological test is based on a basic hypothesis: the ratio  between baryons and total matter (baryons plus dark matter) in galaxy clusters is a fair sample for the Universe on large scales, being constant through the cosmic history. 

{  Over the years, this key hypothesis has been supported by observational and hydrodynamical simulation results. For instance, the Ref.\cite{lagana} investigated the baryon distribution in groups and clusters. By considering 123 systems ($0.02 < z < 1$) they found that the gas mass fraction does not depend on the total mass  for systems  more massive than $10^{14}$ solar masses. Moreover, they obtained  only a slight dependence of gas mass fraction measurements with redshift  for $r_{2500}$ (see their fig. 6).  On the other hand, the hydrodynamical simulations of \cite{battaglia,planelles} showed that hot, massive galaxy clusters ($M_{500}>10^{14}$ solar masses) and dynamically relaxed,  do not show  significant evolution for the depletion factor, $\gamma=f_{gas}/(\Omega_b/\Omega_M)$, (see Table III in \cite{planelles}). They considered $\gamma = \gamma_0 +\gamma_1z$ and found $-0.02 < \gamma_1 < 0.07$ considering the complete sphere at $r_{2500}$ (this radii is that one within which the mean cluster density is 2500 times the critical density of the Universe at the cluster's redshift). However, it is  important to comment that thanks to new X-ray observations, it has been possible detect the presence of intrinsic scatter in the gas mass fraction measurements. In 40 measurements from the Ref. \cite{mantz2014}, for instance, a 7.4\% of intrinsic scatter was found. At the moment, it is not possible to distinguish observationally  between the possible causes of this scattering. In this way, hydrodynamical simulations  have also shown  that a similar level of dispersion may be due to presence of a non-thermal pressure (see, for instance, \cite{Nelson}).}

Recently, some  works have shown that the X-ray gas mass fraction measurement as a cosmological tool is strongly dependent on the cosmic distance duality relation (CDDR) validity \cite{gon2012,hga2012,santos,wang}, $D_LD_A^{-1}=(1+z)^{2}$, where $D_L$ and $D_A$ are the luminosity and angular diameter distances for a given redshift $z$. {  Particularly, the authors of Ref.\cite{shafeliou} searched for systematics in SNe Ia and galaxy cluster data using this relation, without advancing any hypothesis about the nature of dark energy}. This relation was proved  {in} Ref. \cite{ethe} and it only requires  sources and observers  connected by null geodesics in a general Riemannian spacetime as well as conservation of photon number\footnote{For recent results of CDDR tests see Table I in \cite{hba2016}. }. Thus, even in a Riemannian spacetime, any departure from cosmic transparency  could lead to dubious estimates of cosmological parameters  if one uses cosmological tests dependent on flux, such as, SNe Ia distance module and  X-ray gas mass fraction (X-ray GMF). Therefore, although  the dark energy is supported by several other independent probes, if some extra opacity is still present, the observations will give us unreal values to cosmological parameters, mainly to the $\Omega_M$ if one considers those from  X-ray GMF of galaxy clusters.  

In this work we discuss how X-ray GMF observations of galaxy clusters jointly with recent $H(z)$ data can be used to investigate a possible departure from transparency cosmic in a flat $\Lambda$CDM framework. The X-ray GMF samples used  separately in our analyses consist of: 42 and 40 measurements obtained by the Refs.\cite{allen2008} and \cite{mantz2014}, respectively. The total redshift range is $0.078 \leq z \leq 1.063$. The  $H(z)$ data consist of 38 points in the redshift range $0.07 \leq z \leq 2.36$ obtained from cosmic chronometers and radial BAO methods (see section III  for details). We also consider a gaussian prior on the Hubble constant value, $H_0$, from Ref.\cite{Ade} ({\it Planck collaboration}). In our analyses, the cosmic opacity is parameterized by $\tau(z)=2\epsilon z$, which corresponds to a modification on the cosmic distance duality relations such as $D_LD_A^{-1}=(1+z)^{2+\epsilon}$ (if $\epsilon$ is small and $z \leq 1$). Our results are consistent with a transparent universe within $1\sigma$ c.l. ($\epsilon \approx 0$).  Although the limits presented here on the cosmic opacity are less restrictive than those coming from the similar analyses with SNe Ia, they correspond to another band of the electromagnetic spectrum. Moreover, we verify that the constraints on $\Omega_M$ obtained from the X-ray GMF test are strongly dependent on  {  cosmic transparency assumption even in the simple  flat $\Lambda$CDM framework\footnote{It is worth to comment that the authors of Ref.\cite{santos} tested the CDDR with gas mass fraction and $H(z)$ measurements in a cosmological model independent approach. In this way, no information of how cosmological results from X-ray observations are depend on the cosmic transparency hypothesis was obtained.}.}

This paper is organized as follows. In Section~\ref{sec:observations} we present the method. The cosmological data are described in Section~\ref{sec:data}. The analyses and results are presented in Section~\ref{sec:results} and Section~\ref{sec:conclusions} shows our conclusions. 

\section{Cosmic opacity and gas mass fraction observations}
\label{sec:observations}

In this section we first discuss how a cosmic opacity presence affects the luminosity distance, and then we present the link between luminosity distance and X-ray GMF observations.

\subsection{Luminosity distance and cosmic opacity}

The methodology used in our analyses was proposed by \cite{avg2009}. It was initially applied for SNe Ia data, however, it also can be applied for X-ray GMF observations.  As it is well known, the distance modulus derived from SNe Ia or gamma-ray bursts and the X-ray surface brightness observations may be  systematically affected if there are  cosmic dimming sources. In few words, a direct consequence of  the photon number reduction is  an increasing of  $D_L$.  {Hence},  if $\tau(z)$ denotes the opacity  between an observer at $z=0$ and a source at $z$, the flux received by the observer in $z=0$  {is} attenuated by a factor $e^{-\tau(z)}$  {and, therefore,} the observed luminosity distance ($D_{L, obs}$) is related to the true luminosity 
distance ($D_{L, true}$) by
 \begin{equation}
 D_{L, obs}^2=D_{L,true}^2 e^{\tau(z)} \, .
 \end{equation}
Then, the \emph{observed} distance modulus is  \cite{chen1,chen2}
\begin{equation}
\label{distancemod}
m_{obs}(z)=m_{true}(z)+2.5(\log e) \tau(z) \, .
\end{equation}
In this paper, $D_{L,true}(z)$ comes from a flat $\Lambda$CDM model, such as
\begin{equation}
\label{dl}
 D_{L,true}(z, \Omega_M, H_0)=(1+z)c\int_o^z\frac{dz'}{H(z)},
 \end{equation}
 where $c$ is the speed of light and
  \begin{equation*}
	\label{hz}
  H(z)=H_0 E(z, \textbf{p})\\, 
  \end{equation*}
  \begin{equation}
  E(z,\textbf{p})=[\Omega_M(1+z)^3+(1-\Omega_M)]^{1/2}.
  \end{equation}
In the above expression, $\Omega_M$ stands for the matter density parameter measured today. In order to use the full redshift range of the available data, we follow the Refs.\cite{avg2009,avg2010} and considered the   parameterization $D_L=D_A(1+z)^{(2+\epsilon)}$, with $\epsilon$ parameterizing departures from transparency cosmic. These authors argued that for small $\epsilon$ and $z \leq 1$ this is equivalent to assume an optical depth parameterization  $\tau(z)=2\epsilon z$ or  $\tau=(1+z)^{\alpha}-1$ with the correspondence $\alpha=2 \epsilon$. In this way, in our analyses, we consider a simple linear parameterization for $\tau(z)$, such as: $\tau(z)=2\epsilon z$. The measurements of $m_{obs}$ (or $D_{L, obs}$) are obtained from the X-ray GMF data. The unknown parameters $\Omega_M$ and $\epsilon$ are constrained by fitting the X-ray GMF data separately and jointly  with $H(z)$ measurements on a flat $\Lambda$CDM model. As comment earlier, the basic idea behind this test is that while the X-ray surface brightness of galaxy clusters can be affected by cosmic opacity, the $H(z)$ measurements  are obtained via two ways: from measurements of radial BAO and from the differential ages of old passively evolving galaxies, which relies only on the detailed shape of the galaxy spectra but not on the  galaxy luminosity. Both methods are cosmic opacity independent. Therefore, $H(z)$ values are not affected by a non-zero  $\tau(z)$ since $\tau$ is assumed  not to be  strongly wavelength dependent in the optical band. 

\begin{figure*}[htb]
\centering
\includegraphics[width=0.47\textwidth]{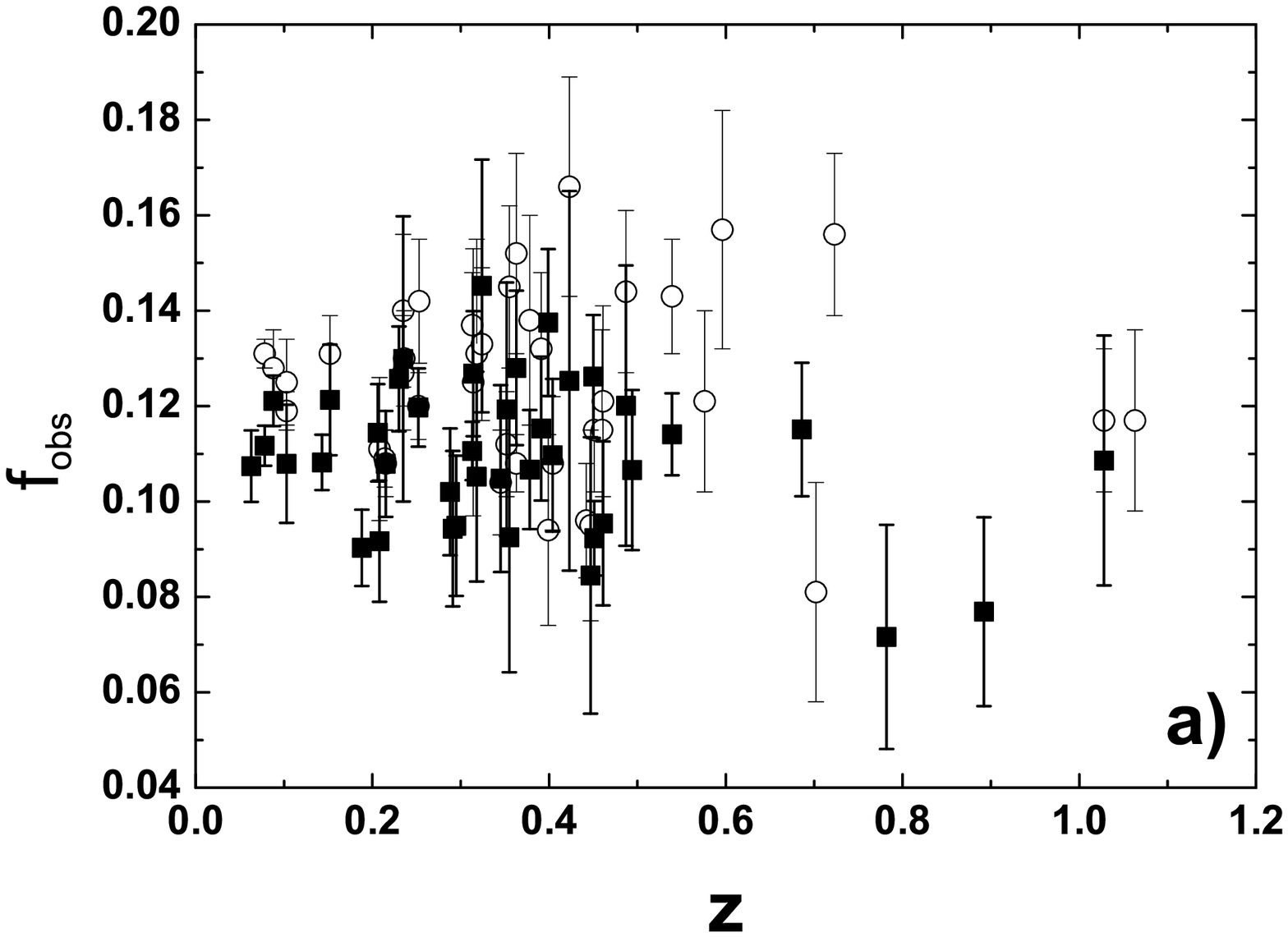}
\hspace{0.3cm}
\includegraphics[width=0.47\textwidth]{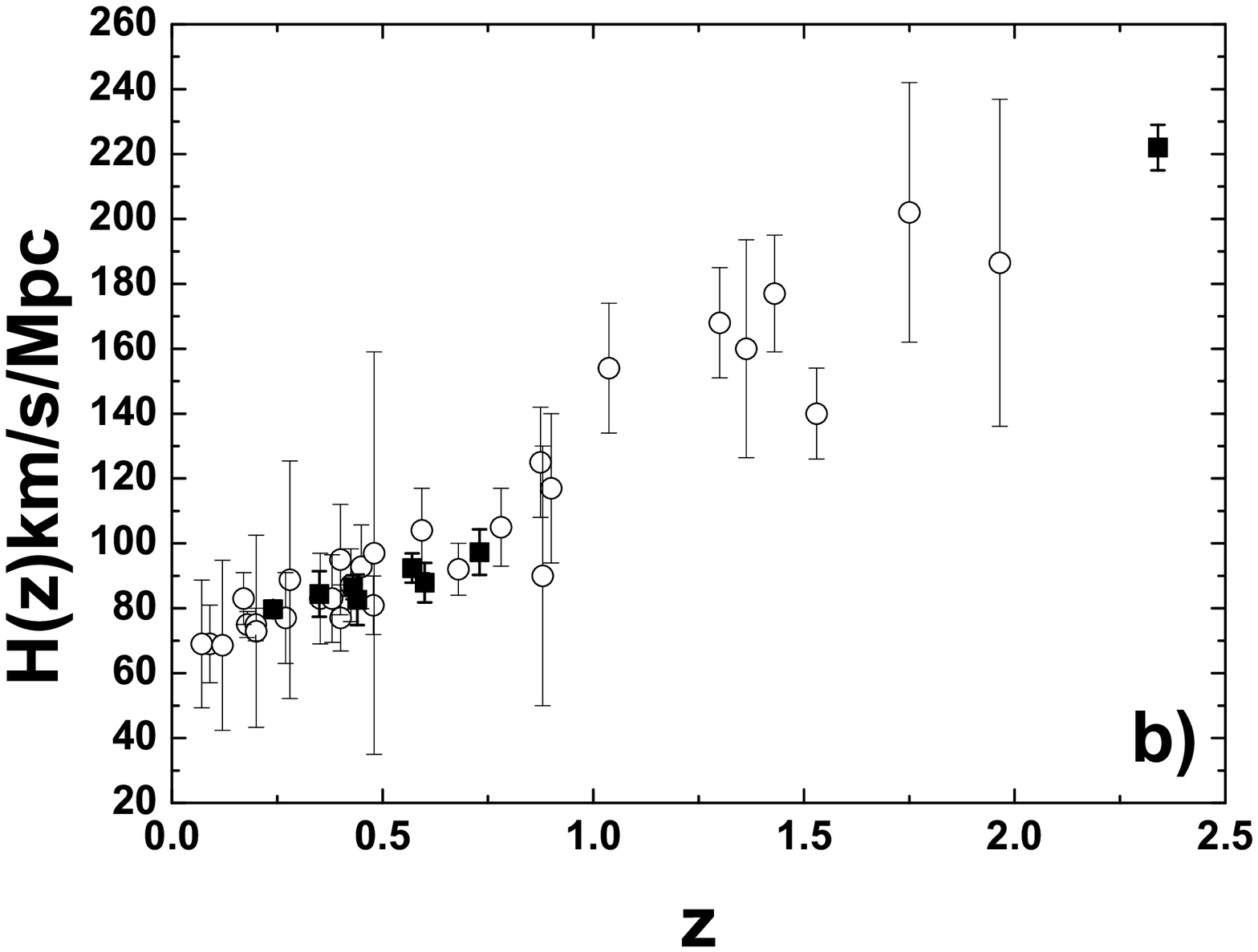} 
\caption{Fig.(a) shows the X-ray GFM data. The open circles and filled squares correspond to samples from Ref.\cite{mantz2014} and Ref.\cite{allen2008}, respectively. Fig.(b) shows the $H(z)$ data (in units of $km/s/Mpc$). The open circles and filled squares correspond to measurements from cosmic chronometers and radial BAO, respectively. }
\end{figure*}
\subsection{Luminosity distance from galaxy clusters}

 In galaxy clusters, the GMF is defined by \cite{allen2008}

\begin{equation}
 f_{gas}=\frac{M_{gas}}{M_{tot}},
  \label{eq3.14}
\end{equation}
where $M_{tot}$ is the total mass (dominated by dark matter)  and  $M_{gas}$ is the gas mass. The total mass within a given radius $R$  can be obtained by assuming that the intracluster gas is in hydrostatic equilibrium. On the other hand,  the intracluster gas  emits X-ray predominantly via  thermal bremsstrahlung and its mass can be estimated by  integrating a gas density model. The $f_{gas}$ is expected to be same at all $z$ since these structures are the largest virialized objects in the Universe, consequently, a faithful representation of the cosmological average baryon fraction can be found in clusters.  Thus, in order to constrain cosmological parameters, the X-ray GMF of galaxy clusters can  be used via the following expression \cite{allen2008}
\begin{equation}
\label{GasFrac}
f^{obs}_{X-ray}(z)=N\left[\frac{D_L^* D_A^{*1/2}}{D_L D_A^{1/2}}\right],
\end{equation}
where the symbol * denotes quantities from a fiducial cosmological model  used in the observations (usually a flat $\Lambda$CDM model with $\Omega_m=0.3$ and $H_0=70$ km/s/Mpc). The parameter $N$ defines an arbitrary normalization on which we marginalize. The ratio multiplying  in brackets computes the expected measured gas fraction $f^{obs}_{X-ray}$ when the cosmology is varied. On the other hand, in Ref.\cite{gon2012}, the authors showed that the gas mass fraction measurements  extracted from X-ray data are affected if there are cosmic opacity sources (consequently, departure from the CDDR validity). In a such framework, if one considers our parametrization $\tau(z)=2\epsilon z$ (or $D_L=D_A(1+z)^{(2+\epsilon)}$), the Eq.(\ref{GasFrac}) is rewritten as
\begin{eqnarray}
\label{GasFrac3}
f^{obs}_{X-ray}(z) &=& N \left[\frac{(1+z)^{\epsilon/2}D_L^{*3/2}}{D_{L}^{3/2}}\right].
\end{eqnarray}
Finally, we define the distance modulus of a galaxy cluster as
\begin{equation}
m_{obs}(z, N, \epsilon)=5\log[(1+z)^{\epsilon/3}D_L^{*}[N/f^{obs}_{X-ray}(z)]^{2/3}]+25, 
\end{equation} 
which depends on cosmic opacity ($D_L^*$ is in Mpc).
 
\begin{figure*}[htb]
\centering
\includegraphics[width=0.47\textwidth]{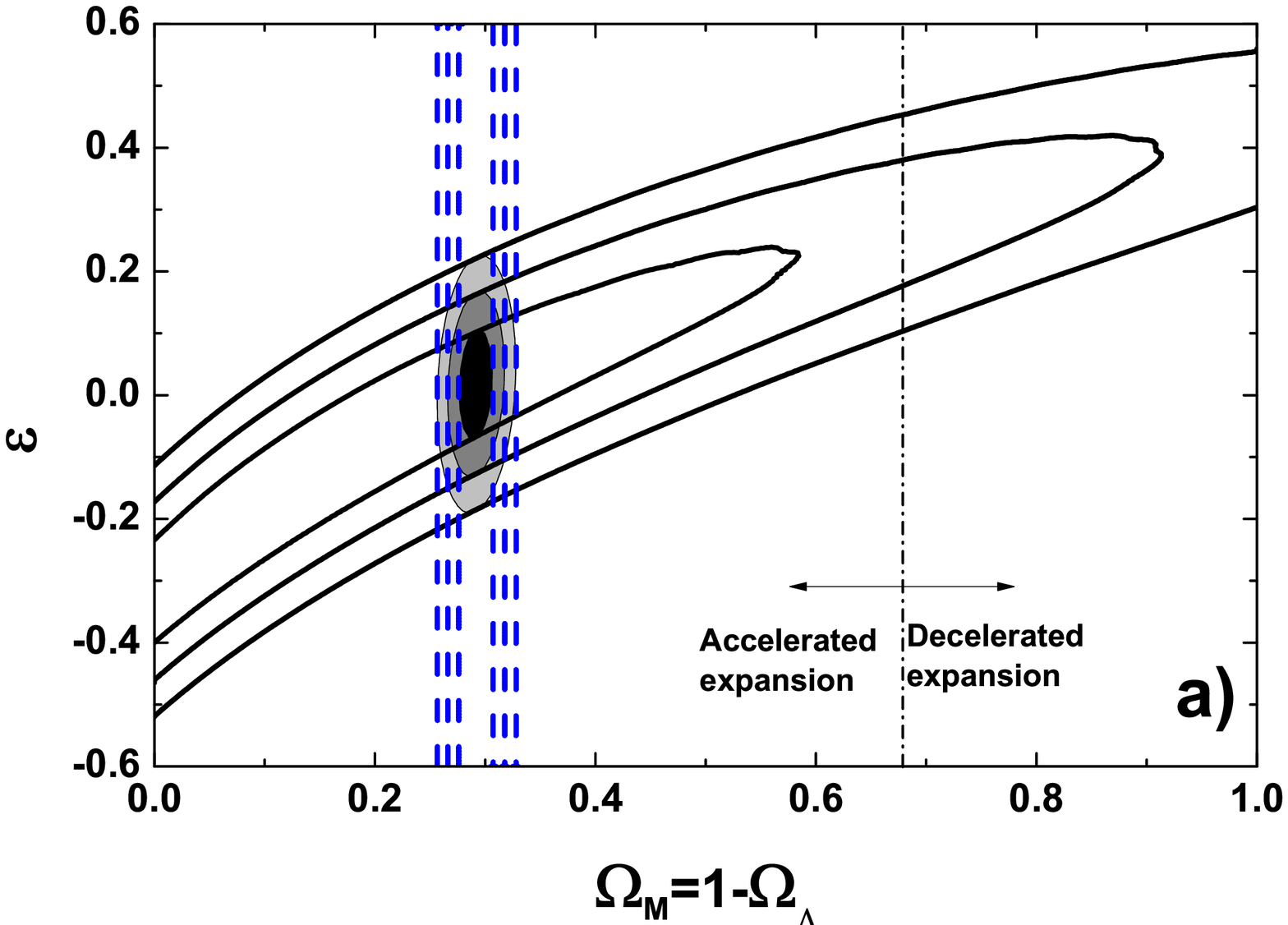}
\hspace{0.3cm}
\includegraphics[width=0.47\textwidth]{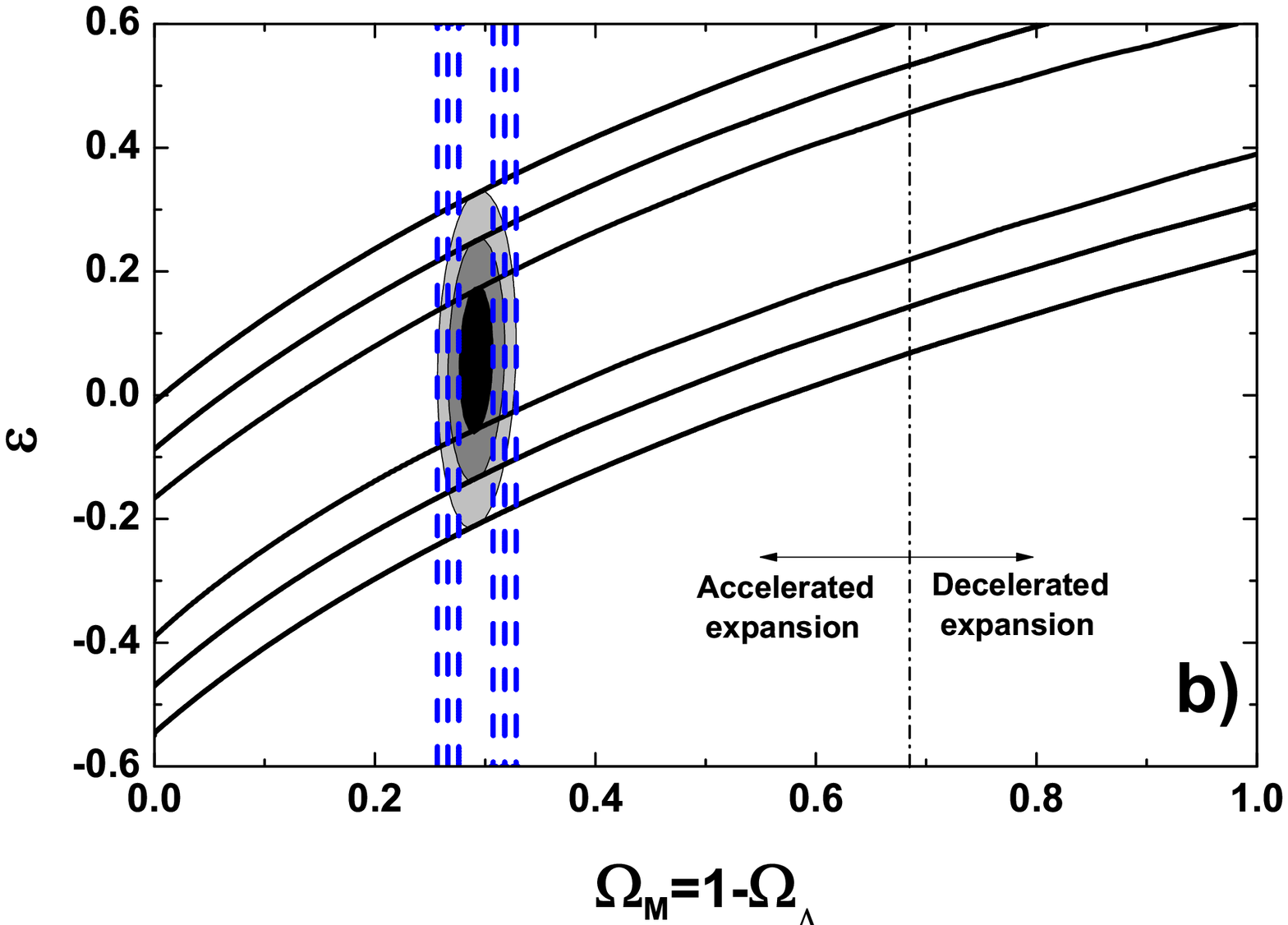}
\\
\includegraphics[width=0.47\textwidth]{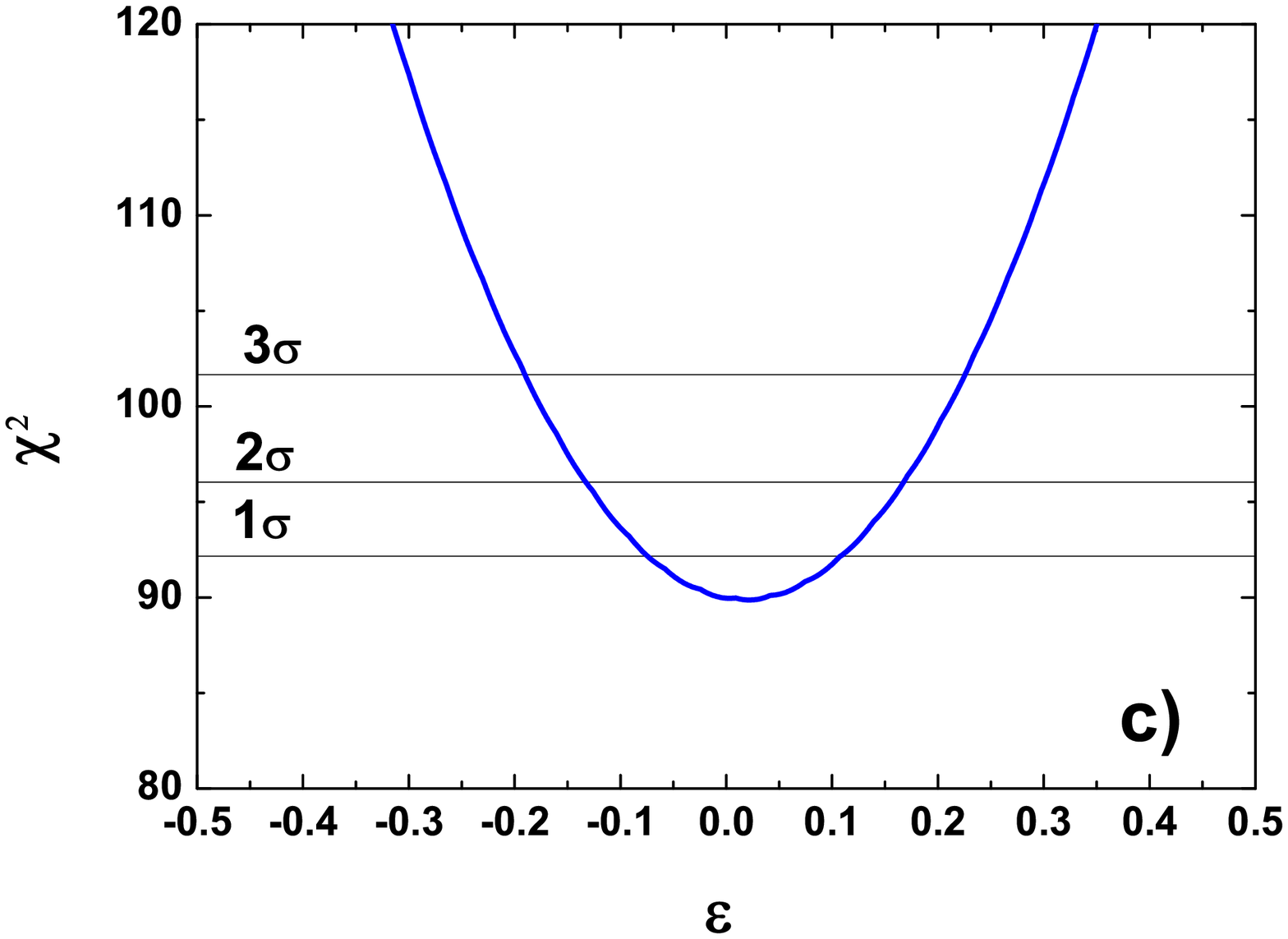}
\hspace{0.3cm}
\includegraphics[width=0.47\textwidth]{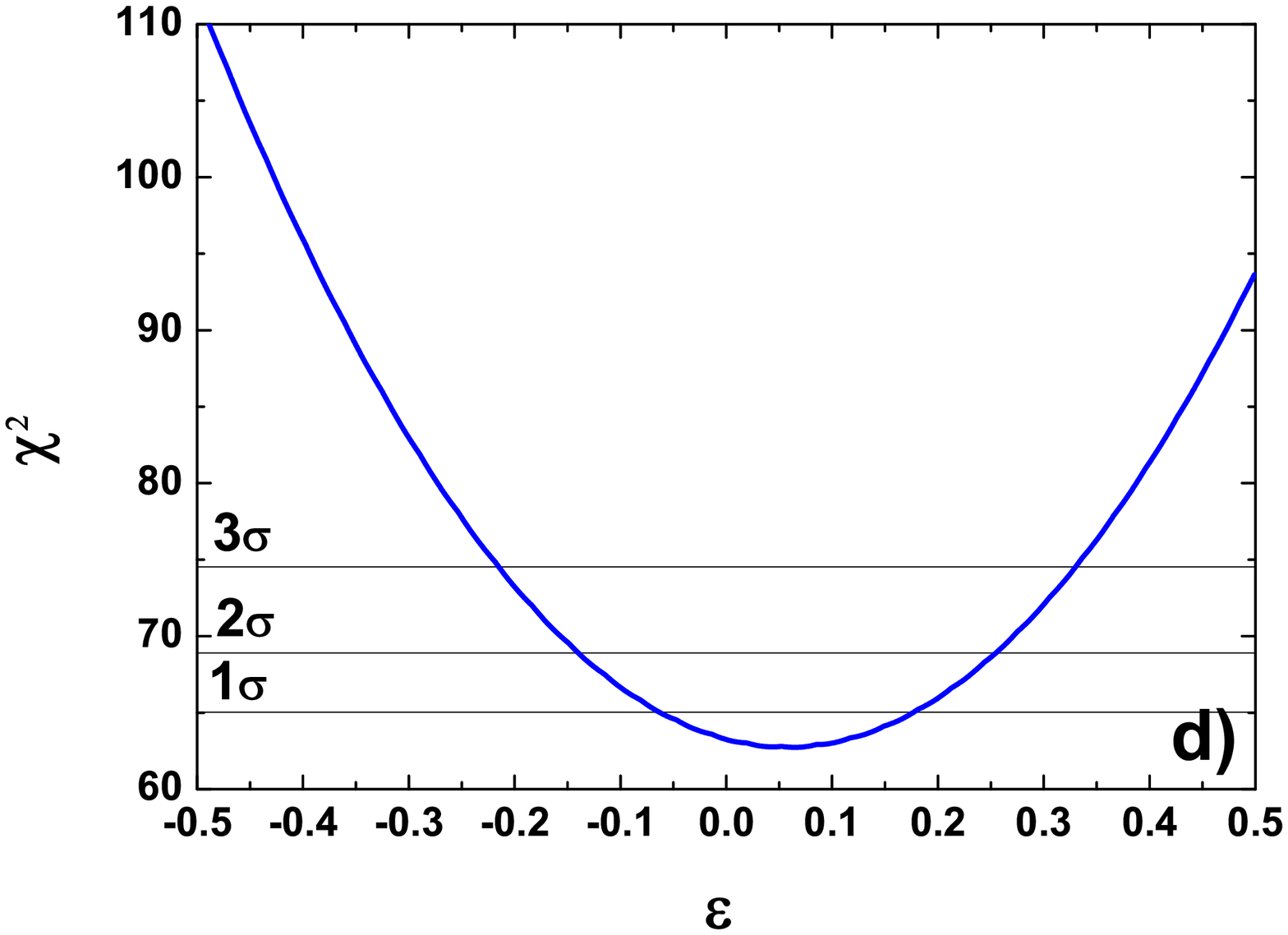}
\caption{{  Fig.(a) shows the  ($\epsilon, \Omega_M$) plane by the using X-ray GMF sample from Ref.\cite{mantz2014} and the $H(z)$ data. Fig.(b) shows the X-ray GMF sample from Ref.\cite{allen2008} and the $H(z)$ data. In all these panels the black and blue lines correspond to analyses by using X-ray GMF samples and $H(z)$ data separately. In each case, the filled contours are the results from joint analysis. Panels (c) and (d) show  the $\chi^2$ values for $\epsilon$ by using the X-ray GMF samples  from Ref.\cite{mantz2014} and Ref.\cite{allen2008}, respectively, jointly with $H(z)$ data (marginalizing on the $\Omega_M$).}}
\end{figure*}

\section{Data}\label{sec:data}

In this paper, we consider two types of data sets:

\subsection{Cosmic opacity dependent data}

Here we have two GMF samples, namely:

\begin{itemize}
\item 42  X-ray GMF measurements obtained by the Chandra telescope for hot ($kT>5keV$), massive, X-ray luminous and dynamically relaxed galaxy clusters spanning the redshift range $0.05 \leq z \leq 1.1$ (see Ref.\cite{allen2008}). The X-ray emitting gas mass and total mass were obtained via  hydrostatic equilibrium and spherical symmetry assumptions. Particularly, the total mass distribution was described by the so-called Navarro, Frank and White profile \cite{navarro}. The measurements for each cluster were performed within $r_{2500}$ radius in the reference $\Lambda$CDM cosmology. This radius corresponds that one for which the mean enclosed mass density is 2500 times the critical density of the Universe at the redshift of the cluster (see Fig.1a).

\item 40 X-ray GMF measurements from massive, dynamically relaxed galaxy clusters compiled by the Ref.\cite{mantz2014}. These authors significantly reduce systematic uncertainties compared to previous papers by  incorporating a robust gravitational lensing calibration of the X-ray mass estimates. Moreover, as an unprecedented  approach, the GMF  measurements were obtained in spherical shells at radii near $r_{2500}$,
rather than X-ray GMF integrated at all radii $< r_{2500}$. This procedure excludes  cluster centers  and reduces the  theoretical uncertainty in gas depletion from hydrodynamic simulations. As a result, the error bars of this sample are smaller than those in Ref.\cite{allen2008} (see Fig.1a).
\end{itemize}

\subsection{Cosmic opacity independent data}

Here, we consider 38 $H(z)$ measurements, namely, 30 from  cosmic chronometers (see table 4 in Ref.\cite{moresco2016}) plus 8 $H(z)$ measurements from radial baryon acoustic oscillations (see Fig.1b). Briefly, the cosmic chronometers approach  uses relative ages of the most massive and passively evolving galaxies to measure $dz/dt$, from which $H(z)$ is inferred. The method of getting ages of old passively evolving galaxies depends only on the detailed shape of the galaxy spectra but not on the  galaxy luminosity, which turns this quantity independent on cosmic opacity\footnote{ We consider $\tau$  not to be strongly wavelength dependent in the optical band (see Refs.\cite{22,23}).}.  The 8 $H(z)$ measurements from radial baryon acoustic oscillations can be found in Refs.\cite{gas,xu,blake,samushia,delu}. A complete data table also can be found in Ref.\cite{shu}. Recently these measurements of the Hubble parameter have been used to constrain several cosmological parameters \cite{simon,stern,chen2011,seik,moresco,Farooq,moresco2016,moresco2016b}. {{We considered in our analyses $H_0 = 67.8 \pm 0.9$, (in $km/s/Mpc$), obtained by the {\it Planck collaboration} for a flat $\Lambda$CDM universe from a combination of temperature and lensing data of the cosmic microwave background  \cite{Ade}}}. 

\begin{table*}[htb]
\caption{Constraints on $\epsilon$ from different analyses. The symbols * and ** denote the samples from Refs.\cite{mantz2014} and \cite{allen2008}, respectively.}
{\begin{tabular} {c||c||c|c|c}
Reference & Data set & Model & $\tau(z)$&$\epsilon$ ($1\sigma$) 
 \\
\hline \hline 
\cite{avg2009}& 307 SNe Ia + 10 $H(z)$ & flat $\Lambda$CDM &  $\tau(z)=2\epsilon z$ &$-0.01^{+0.06}_{-0.04}$ \\
\cite{avg2010}& 307 SNe Ia + 12 $H(z)$&flat $\Lambda$CDM&  $\tau(z)=2\epsilon z$ &$-0.04^{+0.04}_{-0.03}$  \\
\cite{HCA2013} &    581 SNe Ia + 28 $H(z)$ & model independent&  $\tau(z)=2\epsilon z$&$0.017 \pm 0.052$ \\
\cite{HB2014}& 581 SNe Ia + 19 $H(z)$ & flat $\Lambda$CDM &  $\tau(z)=\epsilon z$&$0.02 \pm 0.055$\\
\cite{HB2014}& 59 GRB +   19 $H(z)$ & flat $\Lambda$CDM & $\tau(z)=\epsilon z$ &$0.06 \pm 0.18$\\
\cite{HB2014}& 581 SNe Ia +19 $H(z)$ & flat XCDM & $\tau(z)=\epsilon z$ &$0.015 \pm 0.060$\\
\cite{HB2014}& 59 GRB + 19 $H(z)$ &flat XCDM &  $\tau(z)=\epsilon z$&$0.057 \pm 0.21$ \\
\cite{Avg2016}& 740 SNe Ia + 19 $H(z)$ & model independent & $\tau(z)=2\epsilon z$& $0.044^{0.078}_{0.080}$\\
 {This paper}& 40 $GMF^{*\dagger}$ + 38 $H(z)$ &flat $\Lambda$CDM &  $\tau(z)=2\epsilon z$&$0.03\pm 0.08$\\
 {This paper}& 42 $GMF^{**\dagger}$ + 38 $H(z)$ &flat $\Lambda$CDM &  $\tau(z)=2\epsilon z$&$0.05 \pm 0.13$\\
\hline
\end{tabular}} \label{ta2}
\end{table*}

\section{Analyses and results}\label{sec:results}

We obtain the constraints to the set of parameters $(\epsilon, N, \Omega_M, H_0)$,  by evaluating the likelihood distribution function, ${\cal{L}} \propto e^{-\chi^{2}/2}$, with

\begin{eqnarray}
\chi^{2} = & \sum_{z}\frac{[m_{obs}(z, N, \epsilon) - m_{true}(z, \Omega_M, H_0)-2.17147\epsilon z]^2}{\sigma^2_{m_{obs}}} \\ \nonumber
                                     &+ \sum_{z}\frac{[H(z, \Omega_M, H_0)- H_{obs}(z)]^2}{\sigma_{H_{obs}}^2} \\ \nonumber
																		 &+ \frac{(H_0- H_0^*)^2}{\sigma_{H_{0}^*}^2}
\end{eqnarray}
where  $\sigma^2_{m_{obs}}$, $\sigma_{H_{obs}}^2$ and $\sigma_{H_{0}^*}^2$ are the errors associated to $m_{obs}(z, N, \epsilon)$ of the galaxy cluster data,  $H(z)_{obs}$ measurements and $H_0$ prior ($H_0 = 67.8 \pm 0.9$, in $km/s/Mpc$), respectively. $m_{true}(z, \Omega_M, H_0)$ is obtained via $m_{true}(z, \Omega_M, H_0)=5\log_{10} D_{L,true}(z, \Omega_M, H_0) + 25$, while $D_{L,true}(z, \Omega_M, H_0)$ is given by equation Eq.(\ref{dl}) and $H(z, \Omega_M, H_0)$ from Eq.(4). We marginalize on the $N$ parameter.

The Fig.(2) shows all the results from our analyses. The black solid contours  in the Figs. (2a) and (2b) are the confidence intervals of $\Delta \chi^2 = 2.30$ $(1\sigma)$, 6.17 $(2\sigma)$ and 11.82 $(3\sigma)$ on the ($\Omega_M -\epsilon$) plane from  analyses with the X-ray GMF samples present in the Refs. \cite{mantz2014} and  \cite{allen2008}, respectively. From these results, it is very important to point out that, even considering the simple flat $\Lambda$CDM model, the constraints on the $\Omega_M$ parameter exclusively from X-ray GMF data  depend  strongly on  the transparency cosmic assumption. This means that using only this kind of observation we can not constrain simultaneously the energy content of the flat $\Lambda$CDM model and the $\epsilon$ parameter. In other words, there is a  degeneracy between the $\Omega_M$ and $\epsilon$ parameters. {  Moreover,  a decelerated universe is allowed  within $\approx$ 1.5$\sigma$ in Fig.(a) and within 1$\sigma$ in Fib.(b) (see the vertical black dashed-dot line)}. In both figures, the vertical blue lines correspond to results by using exclusively the $H(z)$ data (the confidence intervals are for $1\sigma$, 2$\sigma$ and 3$\sigma$). As one may see, the $\Omega_M$ parameter is well constrained when the $H(z)$ data are added in the analyses and, therefore, limits on $\epsilon$ can be found. In each figure, the results from the joint analysis by using X-ray GMF + $H(z)$ are displayed by the filled contours. 

On the other hand, Figs. (2c) and (2d) show  the $\chi^2$ values for $\epsilon$ by using the X-ray GMF samples  from Ref.\cite{mantz2014} and Ref.\cite{allen2008}, respectively, jointly with $H(z)$ data (marginalizing on $\Omega_M$).
The intervals found are (at $1\sigma$): 

\begin{itemize}
\item Fig.(2a): $\Omega_M=0.29 \pm 0.02$ and  $\epsilon=0.03\pm 0.09$.
\item Fig.(2b): $\Omega_M=0.30 \pm 0.02$ and  $\epsilon=0.05\pm 0.14$.
\item Fig.(2c): $\epsilon=0.030 \pm 0.080$ (by marginalizing on $\Omega_M$). 
\item Fig.(2d): $\epsilon=0.05 \pm 0.13$ (by marginalizing on $\Omega_M$). 
\end{itemize}
As one may see, although in this case we have $\epsilon>0$, the results are in full agreement with a transparent universe ($\epsilon=0$). {  Our $\Omega_M$ value is in full agreement with that one from Planck results \cite{Ade}.}

Table~\ref{ta2} shows some recent constraints on the cosmic opacity by using approaches involving SNe Ia, gamma-ray bursts and $H(z)$ observations as well as the results of the present paper. {  As commented earlier, the approach used in Refs.\cite{avg2009,avg2010,HB2014} is similar to this paper, but the bands of the electromagnetic spectrum explored were other, namely, optical and gamma-ray. On the other hand, the Refs.\cite{HCA2013,Avg2016} considered cosmological model independent approaches by using SNe Ia and $H(z)$ data. As one may see, the results from different bands of the electromagnetic spectrum are in full agreement each other and no significant deviation from transparent universe is verified. However, these results do not  rule out $\epsilon \neq 0$ with high statistical significance yet.}

\section{Conclusions}\label{sec:conclusions}

As it is largely known, cosmic opacity can mimic a dark energy behavior and its presence has been investigated along the years by different methods. Recently, type Ia supernovae and gamma ray bursts observations have been used  along with cosmic expansion rate measurements, $H(z)$,  to constrain possible departures from cosmic transparency. In this context, dependent and independent cosmological model analyses were performed. In this paper, by considering a flat $\Lambda$CDM framework we showed how is possible use galaxy cluster X-ray gas mass fraction samples jointly with the most recent $H(z)$ data to impose limits on cosmic opacity. { {We considered the $H_0$ prior in our analyses, namely:  $H_0 = 67.8 \pm 0.9$ obtained by the Planck collaboration \cite{Ade}.}}  

Our results can be found in Fig.(2) and Table I, where $\tau(z)$ quantifies the cosmic opacity and was parameterized by $\tau(z)=2\epsilon z$ in our case. This kind of $\tau(z)$ function is directly linked to a violation of the cosmic distance duality relation validity such as $D_LD_A^{-1}=(1+z)^{2+\epsilon}$ if $\epsilon$ is small and $z \leq 1$. As one may see from  {Table~\ref{ta2}}, we did not find any significant departure from cosmic transparency ($\epsilon \approx 0$) and our results are in full agreement with  previous studies where type Ia supernovae and gamma-ray burst observations were used in similar approaches. However, it is very important to stress that these analyses did not  rule out $\epsilon \neq 0$ with high statistical confidence level and additional tests are still required with forthcoming data. Moreover, from panels (a) and (b) in Fig.(2) it is possible to conclude that constraints on $\Omega_M$ obtained from X-ray gas mass fraction test (black solid contours) depend strongly on the cosmic transparency hypothesis even if the simple flat $\Lambda$CDM model is considered.

\begin{acknowledgements}
RFLH acknowledges financial support from  CNPq  (No. 303734/2014-0).  VCB is supported by S\~ao Paulo Research Foundation (FAPESP)/CAPES agreement under grant 2014/21098-1 and S\~ao Paulo Research Foundation under grant 2016/17271-5. KVRAS is supported by CAPES.
\end{acknowledgements}

\end{document}